\documentclass{PoS}
\usepackage{slashed}

\title{Implementation of the twisted mass fermion operator in the QUDA library}

\ShortTitle{Implementation of the twisted mass fermion operator in the QUDA library}

\author{\speaker{Alexei Strelchenko}\\
Scientific Computing Division, Fermilab, Batavia, IL 60510-5011, USA \\
E-mail: \email{astrel@fnal.gov}
}

\author{Constantia  Alexandrou\\
Department of Physics, University of Cyprus, P.O. Box 20537, 1678 Nicosia, Cyprus, and\\  
Computation-based Science and Technology Research  
Center, Cyprus Institute, 20 Kavafi Str., Nicosia 2121, Cyprus \\  
E-mail: \email{alexand@ucy.ac.cy}
}

\author{Giannis Koutsou\\ 
Computation-based Science and Technology Research Center, Cyprus Institute, 20 Kavafi Str., Nicosia 2121, Cyprus\\ 
E-mail: \email{g.koutsou@cyi.ac.cy}
}
    
\author{Alejandro Vaquero Avil\'es-Casco\\ 
Computation-based Science and Technology Research Center, Cyprus Institute, 20 Kavafi Str., Nicosia 2121, Cyprus\\ 
E-mail: \email{a.vaquero@cyi.ac.cy}
}
    
\abstract{We discuss an  extension of  the QUDA library for the Wilson twisted mass operator. A performance analysis is presented for both degenerate and non-degenerate flavor doublets. The degenerate twisted mass fermion operator runs at up to 190, 487 and 856 Gflops, for double, single and half precisions respectively on recent NVIDIA Kepler GPUs, while our implementation for the non-degenerate flavor doublet allows to reach 163, 516 and 879 GFlops, respectively. The code is currently in production for the hadron structure study.}

\FullConference{31st International Symposium on Lattice Field Theory LATTICE 2013\\
                 July 29 -- August 3, 2013\\
                 Mainz, Germany}

\begin{document}

\section{Introduction}
High performance computing heterogeneous architectures, based  either on GPUs or recently released Intel's MIC co-processors, provide a practical solution to accelerate time consuming lattice QCD (LQCD) tasks \cite{Egri:2006zm}, \cite{Joo:isc2013}. One example of such a task
is the evaluation of disconnected diagrams, or fermion vacuum loops, that have typically been omitted from LQCD calculations due to their large computational cost.
The systematic uncertainties introduced by this omission remains an open issue.

Since accelerators are becoming all the more readily available
components of contemporary and future HPC systems, it is essential to develop multi-platform tools to 
enable LQCD computations to be performed with high efficiency
across a variety of architectures.
The USQCD SciDAC software suit can be considered as a good example of this approach. In particular, this software is made up of software library modules targeting different types of commodity clusters (and custom facilities)  that can be re-used by high-level application packages such as Chroma, CPS or MILC.

In this report we will focus on the twisted mass operator for GPUs, implemented in the QUDA library, a community code based on the CUDA platform for carrying out the time-consuming components of LQCD computations on NVIDIA GPUs \cite{Clark:2009wm}, \cite{Babich:2011np}. There are a number of software packages designed for simulating 
twisted mass LQCD, most notable of which is tmLQCD, originally developed for x86 microarchitectures (though it currently contains a CUDA port as well) \cite{Jansen:2009xp}. An OpenCL implementation is reported in Ref.~\cite{Bach:2012iw}.
	
\section{QUDA programming for twisted mass fermions}
The QUDA library is a GPU code which implements a number of fermion operators (and other helper kernels) and a host interface, in an Object-Oriented Programming paradigm. In particular, each type of Dirac operators as well as host and device spinor fields are encapsulated into separate classes with all the necessary functionality for any third party client code such as, e.g. Chroma, MILC etc. 
  In this section we provide some implementation aspects of the twisted mass code development in QUDA. In the most general case of the mass non-degenerate flavor doublet, the Wilson twisted mass fermion operator formulation reads:\\
\begin{equation}
~~~\slashed{D}_{TM} = \slashed{D}_{W} + i \bar{\mu} \gamma_5 \tau^{3} + \bar{\epsilon} \tau^1,
\end{equation}
where $\slashed{D}_W$ stands for the Wilson term, $\tau^i$ denotes the $i$th $SU(2)$ Pauli matrix and $\bar{\mu}$ and $\bar{\epsilon}$ are the (bare) twisted mass parameters. 
For internal computations QUDA adopts a non-relativistic basis  for  the spinor projections; this allows to
reduce memory traffic while computing hopping terms in time direction. However, the library provides converting routines that can be utilized to transform from a chiral basis into the QUDA internal format: this option is controllable via {\texttt{QudaInvertParam}} interface structure described briefly in Subsection 2.2.

\subsection{Implementation details}
  For the QUDA twisted mass iterative solvers  one can employ two types of (even-odd) preconditioning: symmetric and asymmetric. For instance, one may deal with the following equivalent ('even-even') preconditioned systems:
\begin{equation}
(R_{ee} - \kappa^2 \slashed{D}_{eo} R^{-1}_{oo} \slashed{D}_{oe})\psi_{e} = b_e - \slashed{D}_{eo}R^{-1}_{oo} b_o;~~~~~
\end{equation}
\begin{equation}
(I_{ee} - \kappa^2 R^{-1}_{ee} \slashed{D}_{eo} R^{-1}_{oo} \slashed{D}_{oe})\psi_{e} = R^{-1}_{ee}(b_e - \slashed{D}_{eo}R^{-1}_{oo} b_o);
\end{equation}
where $R$ represents a local twisting operator and the odd component of the solution is reconstructed by the expression:
\begin{equation}
\psi_{o} = R^{-1}_{oo}(b_o - \slashed{D}_{oe}R^{-1}_{ee} \psi_e). 
\end{equation}
Accordingly, we implemented a number of 'fused' CUDA kernels, such as
$~  R^{-1}_{oo} \slashed{D}_{oe}, (R_{ee} - \kappa^2 \slashed{D}_{eo})$ (and their 'daggered' analogues), required for the left-hand-side (LHS) of Eq. (2.2). As a result, all local operators are merged into dslash kernels and computed on the fly reducing expansive accesses to the GPU global buffer. All these kernels are generated by a python script in the same way as it is done for other fermion operators available in QUDA.

Next, the main peculiarity of the non-degenerate twisted mass fermion operator consists of the presence of off-diagonal matrix elements in flavor subspace, introduced by the third term in the right-hand-side (RHS) of Expr. (2.1). That is, in this case, one has to apply the dslash on both flavors resulting in more complicated compute kernels
and the most straightforward approach here is to re-use the gauge field to avoid an extra memory transaction while computing contributions from each spinor flavor.

Finally, to include the twisted mass dslash operator in the whole framework, we added two new classes, {\texttt{DiracTwistedMass}} and {\texttt{DiracTwistedMassPC}}, which encapsulate  all necessary attributes and methods for both degenerate and non-degenerate flavor doublets, including methods for launching dslash kernels on the accelerators. The multi-GPU parallelization for the degenerate flavor doublet is almost identical to the corresponding Wilson implementation. On the contrary, for the non-degenerate case, since matrix-vector operations involve  two fermion flavors, we had to redesign QUDA packing routines to properly take into account the fifth flavor dimension when gathering boundary-spinor sites in non-temporal lattice directions.

More detailed information about optimization strategies exploited in the QUDA library can be found in Refs.~\cite{Clark:2009wm, Babich:2011np, Clark:2012aa}.

\subsection{End-user configuration and setup guide}
To compile the twisted mass component in the QUDA library one needs to provide  the package configure script with a new option, i.e,
\begin{center}
{\textcolor{black}{\texttt{--enable-[ndeg-]twisted-mass-dirac}}.}
\end{center}
The end-user application setup is pretty similar to the Wilson case. Namely,  one should specify the following key information to QUDA by means of {\texttt{QudaInvertParam}} structure attributes declared in {\texttt{quda.h}} header file of the library:\\
  -- dslash type ({\texttt{dslash\_type}} attribute). For the twisted mass operator one can currently choose between {\small \texttt{QUDA\_TWISTED\_MASS\_DIRAC}} or {\small \texttt{QUDA\_NDEG\_TWISTED\_MASS\_DIRAC}}
     for the degenerate or non-degenerate flavor doublets, respectively.\\
  -- flavor degeneracy ({\texttt{twist\_flavor}} attribute). Available options are:  {\small \texttt{QUDA\_TWIST\_PLUS}}, {\small \texttt{QUDA\_TWIST\_MINUS}} or {\small \texttt{QUDA\_TWIST\_NDEG\_DOUBLET}}.\\
  -- the twisted mass parameters $\mu$, $\epsilon$. Note that $\epsilon$ is set to zero by default for the degenerate case.

To setup the QUDA solvers, a client application should also specify {\texttt{solver\_type}} and {\texttt{solution\_type}} attributes of the interface. The former one indicates whether to solve the original ($Ax=b$) or normal ($A^{\dagger} Ax=A^{\dagger}b$) linear system, i.e., {\small \texttt{QUDA\_DIRECT[NORMOP]\_SOLVE}},
and whether the solver has to take care of preconditioning, e.g., {\small \texttt{QUDA\_DIRECT\_PC\_SOLVE}} etc.
In addition to this information, the latter attribute defines whether the (e.g., unpreconditioned) system to be solved has the form $Ax=b$ for {\texttt{DIRECT}} option or $A^{\dagger} Ax=A^{\dagger}b$ for {\texttt{NORMOP}} option ({\small \texttt{QUDA\_MAT\_SOLUTION}}),  or has the form $A^{\dagger}y=b,~Ax=y$ for {\texttt{DIRECT}} option or $A^{\dagger} Ax=b$ for {\texttt{NORMOP}} option ({\small \texttt{QUDA\_MATDAG\_MAT\_SOLUTION}}, respectively).

A complete example of the QUDA interface setup and solvers usage can be found under the \texttt{tests} directory of the package.
\section{Performance analysis}
The twisted mass code was tested on NVIDIA Kepler GPUs based on the recent GK110 micro-architecture. We will analyze both single and multi-GPU performance. The single GPU benchmarks were  preformed on a GTX Titan card that is similar to the Tesla K20X accelerator. For the multi-GPU tests we made use of the K20 cluster at Jefferson Lab. (ECC was enabled on the JLab K20 cluster and disabled on the GTX Titan).

\begin{figure}
\caption{Strong scaling of the Conjugate Gradient algorithm for inverting the flavor-degenerate Twisted Mass fermion operator, using 8-parameter reconstruction of the gauge links. We show results for the double-single (DS, blue circles) and double-half (DH, red circles) mixed precision CG inverter.}
\label{fig:deg}
\begin{center}
  \includegraphics[width=0.8\linewidth]{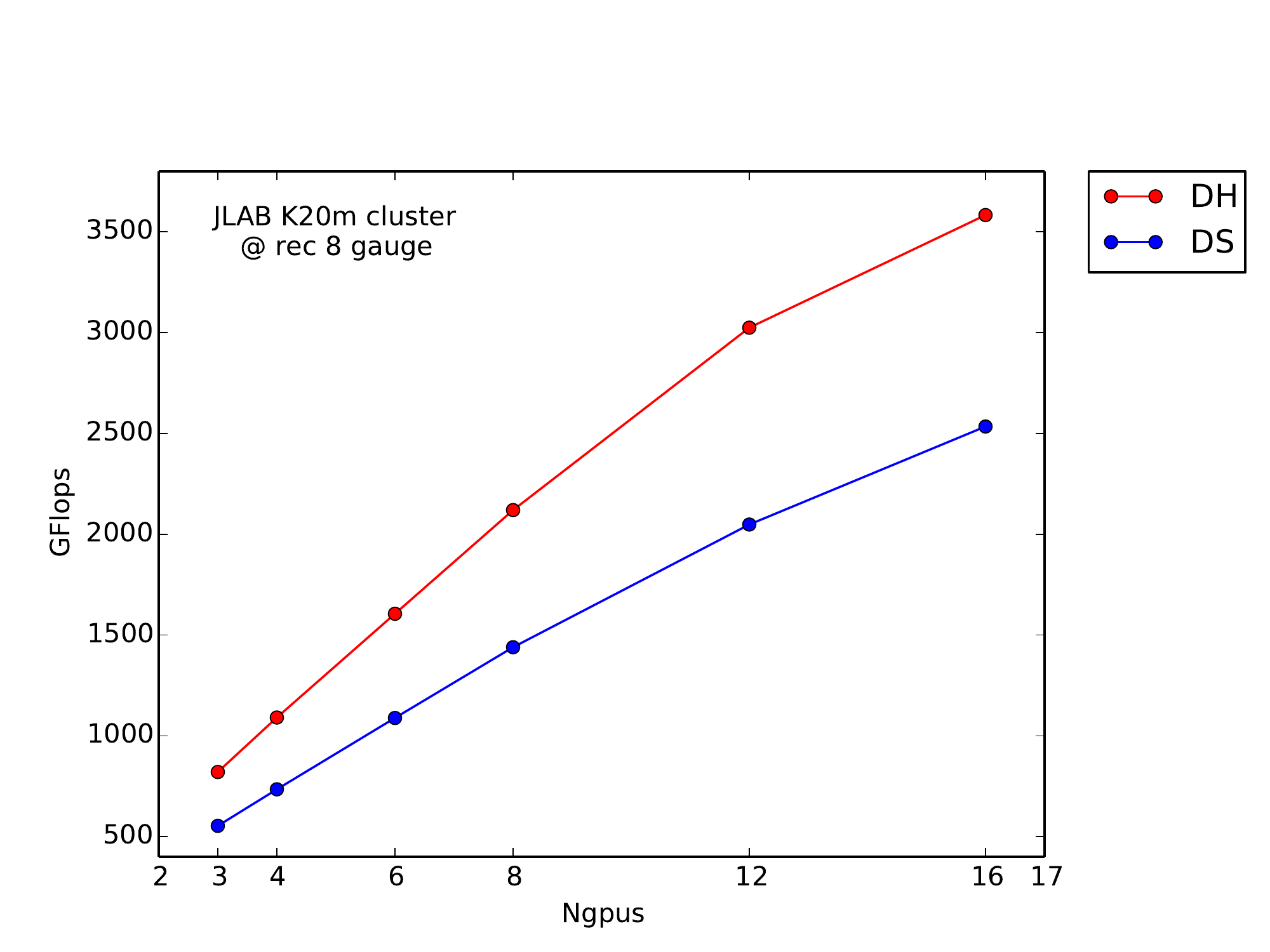}
\end{center}
\end{figure}

We start our analysis with single GPU performance for the (asymmetrically preconditioned) dslash operator, which corresponds to the operator entering the LHS of Eq. (2.2). Here we included the plain Wilson case as a reference point. The lattice size for the single-GPU runs was $32^3 \times 64$ and we examined two types of gauge field reconstructions, namely 8- and 12-parameter reconstructions. QUDA allows for storing the gauge-field links in less than the 9 complex numbers needed to store a full SU(3) matrix. In one case, it allows omitting one row of the three, reducing the storage requirements to 6 complex numbers, so-called 12-parameter reconstruction. With 8-parameter reconstruction, the link is decomposed into a linear combination of the eight SU(3) generators and only the coefficients are stored (8 real numbers). In both cases the full SU(3) is recomputed on the fly during the Dirac operator application. This reduces both the memory requirements but more importantly the bandwidth requirements of applying the Dirac matrix. In addition, to benefit from full-clock speed for the double precision Arithmetic Logic Units on the gaming card we set
\begin{center}
{\texttt{nvidia-setting -a [gpu:0]/GPUDoublePrecisionBoostImmediate=1}}. 
\end{center}
 We summarize our results  in Table 1. 
\begin{table}
\caption{Single GPU performance in GFlops.} \label{tab:single-gpu} 
\begin{center}
\begin{tabular}{|c|c|c|c|c|}
  \hline
  Prec. & Recon. & Wilson & Deg. TM & Non-deg TM\\
  \hline
  \hline
  double~~&~~12& 184~~~ & 190~~~~ & 163~~~~~~ \\
  ~~~~~~~~&~~8& ~179~~~~~ & ~~183~~~~~~ & 115~~~~~~\\
\hline
  single~~~&~~12& 401~~~ & 415~~~~ & 516~~~~~~ \\
  ~~~~~~~~~&~~8& ~472~~~~ & ~~487~~~~~ & 567~~~~~  \\
\hline
  half~~&~~12& 732~~~ & 759~~~~ & 879~~~~~~ \\
  ~~~~~~&~~8& ~829~~~~ & ~~858~~~~~ & 624~~~~~~  \\
  \hline
\end{tabular}
\end{center}
\end{table}

Let us make a few remarks about the obtained results. First, one can observe a consistent degradation in double precision performance for all three fermion operators for the 8-parameter reconstruction, which is due to transcendental operations required for this type of reconstruction. Second, in the non-degenerate case, there is a performance penalty in half precision dslash for the 8-reconstruction, which is attributed to the register spilling.
Taking these observations into account, in the following multi-GPU tests we will consider the 8-parameter reconstruction for the degenerate twisted mass solver and 12-parameter reconstruction for the non-degenerate one.

Our next goal is to illustrate multi-GPU performance profile of the mixed precision degenerate twisted mass CG solver that is combined with asymmetric and symmetric preconditioning. These runs were performed for  $48^3 \times 96$ lattice, with $\kappa = 0.156361, ~~\mu = 0.0015$. Here we provide information on the number of iterations and total solver time depending on number of GPUs and preconditioning type used. For the double-single mixed precision (and the solver tolerance set at $10^{-6}$) the results are presented in Table 2.
\begin{table}
\caption{Multi-GPU performance of the double-single mixed precision CG.} \label{tab:mgpu double-single} 
\begin{center}
\begin{tabular}{|c|c|c|c|}
  \hline
  N GPUs & Asymm. (iter/secs) & Symm. (iter/secs) & Speedup\\
  \hline
  \hline
  4   & \textcolor{red}{3130 / 126.28} & 3131 / 132.72 & \textcolor{red}{~5\%} \\
  6   & \textcolor{red}{3111 / 84.99~} & 3111 / 89.72~ & \textcolor{red}{~6\%} \\
  8   & \textcolor{red}{3169 / 71.93~} & 3169 / 79.47~ & \textcolor{red}{~10\%} \\
  \hline
\end{tabular}
\end{center}
\end{table}
We conclude that the asymmetrically preconditioned CG solver outperforms the symmetrically preconditioned version and the symmetric case requires further optimization.
  
Finally, we present the strong scaling results for the multi-GPU asymmetrically 
preconditioned mixed precision CG solver. We consider here a $32 \times 192$ lattice to demonstrate the best case code scaling. Fig.~\ref{fig:deg} corresponds to the degenerate flavor doublet where we choose 8-parameter reconstruction for the (random) gauge field configuration. 
\begin{figure}
\caption{Strong scaling of the CG algorithm for inverting the flavor non-degenerate Twisted Mass fermion operator, using 12-parameter construction of the gauge links. The rest of the notation is the same as in Fig.~1.}
\label{fig:nondeg}
\begin{center}
  \includegraphics[width=0.8\linewidth]{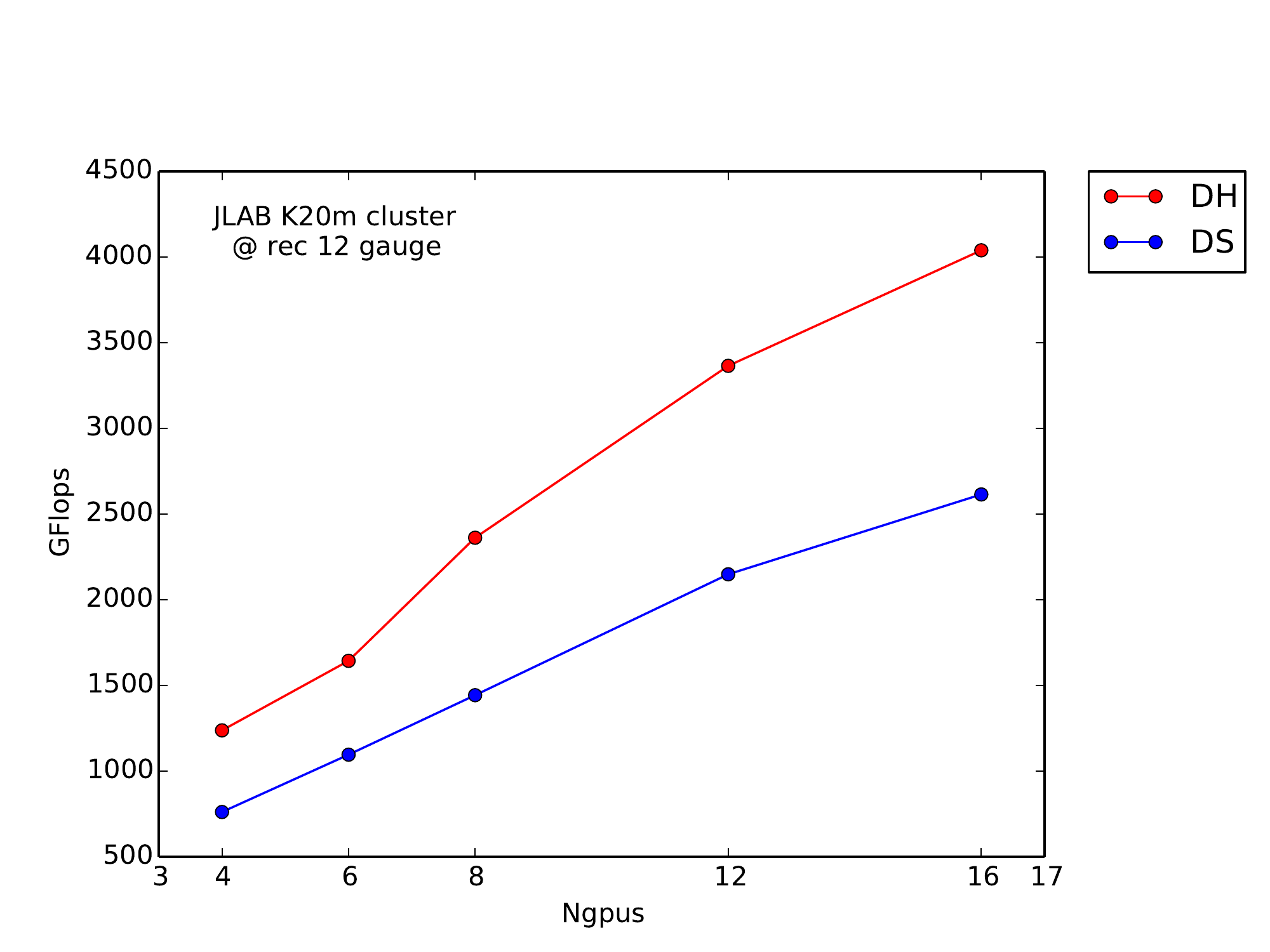}
\end{center}
\end{figure}
Fig.~\ref{fig:nondeg} presents the strong scaling for the non-degenerate case. Here we choose 12-parameter reconstruction as it gives better performance, as can be seen from the comparison in Table~1.
The main difference in the CG performance between degenerate and non-degenerate cases consists in the necessity to invert on both flavors simultaneously in the latter case. As a result, the arithmetic intensity (flop-to-byte ratio) is slightly higher for the non-degenerate flavor doublet, also due to the gauge field re-use mentioned in Section 2.

\section{Conclusion}

The QUDA library was extended to implement an extra fermion operator thus
extending the potential user base of this software package and will allow utilizing NVIDIA accelerators for a
wider set of problems, in particular, problems which are relevant to the European
Twisted Mass collaboration, one of the largest collaboration in Europe. The code is now in production 
and has been used in calculations of disconnected fermion loops such as in Refs.~\cite{Alexandrou:2013wca,Abdel-Rehim:2013wlz}.

Future developments include an implementation, which combines clover-improvement with twisted mass fermions, for the analysis of gauge configurations that have been produced at physical pion mass \cite{Bartosz}. 

\section{Acknowledgements}
A. S. was supported in part by the Research Promotion Foundation of Cyprus under grant $\Pi$PO$\Sigma$E$\Lambda$KY$\Sigma$H/$\Pi$PONE 0308/09, and A. V.  is supported by funding received from the Cyprus Research Promotion Foundation  under contract
EPYAN/0506/08. This work is partly supported by the PRACE-1IP and PRACE-2IP (Community Codes Development - Work Package 8) projects funded by the EUs 7th Framework Programme (FP7/2007-2013) under grant agreement no. RI-211528 and  no. RI-283493 respectively, and by the SciDAC 3 project.
This talk was a part of a coding session sponsored partially by the PRACE-2IP project, as part of the  "Community Codes Development" Work Package 8.

\end{document}